\documentclass[aps,pre,twocolumn,preprintnumbers,floatfix,nofootinbib,10pt]{revtex4-1}
\pdfoutput=1
\usepackage[english]{babel}
\usepackage{amsmath}
\usepackage{array}
\usepackage{geometry}
\usepackage{color}
\usepackage{appendix}
\usepackage{mathtools}
\usepackage{tikz}
\usepackage{multirow}
\usepackage[colorlinks=true,
    linkcolor=blue,
    filecolor=magenta,      
    urlcolor=blue,
    citecolor=blue,
    pdftitle={Sharelatex Example},
    pdfpagemode=FullScreen]{hyperref}

\newcommand{\gmt}{g_{\mu\tau}}
\newcommand{\mzp}{m_{Z'}}
\newcommand{\mdm}{m_\chi}
\newcommand{\mm}{m_\mu}
\newcommand{\mt}{m_\tau}
\newcommand{\lmlt}{L_\mu - L_\tau}
\newcommand{\Lag}{\mathcal{L}}
\newcommand{\Phis}{\Phi_s}

\begin{document} 


\title{\texorpdfstring{Enabling Thermal Dark Matter within the Vanilla $\boldsymbol{L_\mu - L_\tau}$ Model}{Enabling Thermal Dark Matter within the Vanilla Lmu - Ltau Model}}

\author{Nicolás Bernal$^{1}$}
\email{nicolas.bernal@nyu.edu}
\author{Jacinto P. Neto$^{2, 3, 4}$}
\email{jacinto.neto.100@ufrn.edu.br}
\author{Javier Silva-Malpartida$^5$}
\email{javier.silvam@pucp.edu.pe}
\author{Farinaldo S. Queiroz$^{2,3,6,7}$}
\email{farinaldo.queiroz@ufrn.br}

\affiliation{$^1$New York University Abu Dhabi, PO Box 129188, Saadiyat Island, Abu Dhabi, United Arab Emirates}
\affiliation{$^2$Departamento de F\'isica, Universidade Federal do Rio Grande do Norte, 59078-970, Natal, RN, Brasil}
\affiliation{$^3$International Institute of Physics, Federal University of Rio Grande do Norte, Campus Universit\'ario, Lagoa Nova, Natal-RN 59078-970, Brazil} 
\affiliation{$^4$Dipartimento di Scienze Matematiche e Informatiche, Scienze Fisiche e Scienze della Terra, Universita degli Studi di Messina, Viale Ferdinando Stagno d’Alcontres 31, I-98166 Messina, Italy}
\affiliation{$^5$Secci\'on F\'isica, Departamento de Ciencias, Pontificia Universidad Cat\'olica del Per\'u, Apartado 1761, Lima, Peru}
\affiliation{$^6$Millennium Institute for Subatomic Physics at the High-Energy Frontier (SAPHIR) of ANID,\\
Fern\'andez Concha 700, Santiago, Chile}
\affiliation{$^7$Universidad de La Serena, Casilla 554, La Serena, Chile}

\begin{abstract}
Thermal dark matter is a compelling setup that has been probed by a multitude of experiments, mostly in the GeV-TeV mass range. The thermal paradigm in the sub-GeV range is about to experience the same experimental test with the next generation of low-energy accelerators and light dark matter detectors. Motivated by this, we investigate thermal dark matter in the $L_\mu-L_\tau$ and assess how the introduction of a matter-dominated era impacts the parameter that yields the correct relic density. Interestingly, we show that the projected experiments, such as MuSIC, FCC-ee, and LDMX, will probe a large region of the viable parameter space that yields the correct relic density. In the GeV-TeV mass regime, the usual large-scale detectors push the sensitivity. Our work highlights the rich interplay between early-universe dynamics, dark matter phenomenology, and the discovery potential of next-generation experiments.
\end{abstract}

\maketitle

\section{\label{sec:intro} Introduction}
Over the past decades, dual-phase xenon direct detection experiments have pushed the sensitivity frontier for dark matter (DM) interactions close to the so-called neutrino floor. Despite this progress, to date no positive signal has been observed~\cite{XENON:2023cxc, LZ:2024zvo}. These null results have placed stringent constraints on Weakly Interacting Massive Particle (WIMP) scenarios, particularly when the DM has sizable couplings to quarks and electrons. However, such scenarios typically rely on the assumption of standard thermal freeze-out for DM production.

The $\lmlt$ model~\cite{He:1990pn, He:1991qd} has gained much attention in the community because of its simplicity and possibility to accommodate anomalies in flavor physics and the anomalous magnetic moment of the muon ($g-2$)~\cite{Baek:2008nz, Heeck:2011wj, Altmannshofer:2014cfa, Altmannshofer:2015mqa, Crivellin:2015mga, Baek:2015fea, Elahi:2015vzh, Biswas:2016yan, Patra:2016shz, Holst:2021lzm, Huang:2021biu}. However, in recent years, the model has come under significant pressure due to stringent limits set by the NA64$\mu$ fixed-target experiment~\cite{NA64:2024klw}. In parallel, substantial progress in the theoretical prediction of the muon anomalous magnetic moment, driven by both the lattice QCD community and the White Paper 2025, has brought the Standard Model (SM) prediction into closer agreement with the experimental measurement~\cite{Boccaletti:2024guq, Davies:2025pmx, Aliberti:2025beg, Muong-2:2025xyk}. Although these developments have further constrained the parameter space of the vanilla $\lmlt$ scenario, there is still much to explore. For example, this framework can accommodate $H_0$ tension~\cite{Escudero:2019gzq}. Also, a simple extension, such as a leptoquark model with gauged $U(1)_{\lmlt}$ can be used to study flavor anomalies in rare processes involving down-type quark transitions with missing-energy final states $d_i \to d_j\,\nu\,\bar\nu$~\cite{Chen:2023wpb}. 

The vanilla $\lmlt$ model has recently been explored assuming that DM is a Dirac fermion via the standard freeze-out mechanism~\cite{Wang:2025kit}. In this work, we extend this study by covering a larger region of parameter space and considering an early matter-dominated (EMD) era in the cosmological history, which can significantly modify the DM relic abundance and relax the bounds derived under standard cosmology. As we shall discuss in detail, this non-standard cosmological phase opens new windows for viable parameter space that may be probed by upcoming experiments, potentially offering insights into the unknown conditions of the primordial universe. 

Simple extensions of the SM can give rise to EMD phases through scenarios involving long-lived heavy scalar fields~\cite{Giudice:2000ex, Fornengo:2002db, Pallis:2004yy, Gelmini:2006pw, Drees:2006vh, Yaguna:2011ei, Roszkowski:2014lga, Drees:2017iod, Bernal:2018ins, Bernal:2018kcw, Cosme:2020mck, Arcadi:2021doo,Arias:2021rer, Asadi:2021bxp, Bernal:2022wck, Bhattiprolu:2022sdd, Chowdhury:2023jft, Haque:2023yra, Ghosh:2023tyz, Silva-Malpartida:2023yks, Barman:2024mqo, Silva-Malpartida:2024emu, Bernal:2024ndy, Belanger:2024yoj,Banerjee:2024caa, Bernal:2025osg} or ultralight primordial black holes (PBHs) that can act as a matter component~\cite{Green:1999yh, Khlopov:2004tn, Dai:2009hx, Fujita:2014hha, Allahverdi:2017sks, Lennon:2017tqq, Morrison:2018xla, Hooper:2019gtx, Chaudhuri:2020wjo, Masina:2020xhk, Baldes:2020nuv, Gondolo:2020uqv, Bernal:2020kse, Bernal:2020ili, Bernal:2020bjf,  Bernal:2021akf, Cheek:2021odj, Cheek:2021cfe, Bernal:2021yyb, Bernal:2021bbv, Bernal:2022oha, Cheek:2022dbx, Mazde:2022sdx, Cheek:2022mmy, Arcadi:2024tib}. These scenarios alter the Hubble expansion rate and the thermal history of the universe prior to the onset of Big Bang nucleosynthesis (BBN), thereby modifying the freeze-out dynamics and the relic abundance of DM~\cite{Allahverdi:2020bys, Batell:2024dsi}. Such modifications can significantly impact the interpretation of current experimental limits, motivating their inclusion in comprehensive studies of DM.

We begin in Section~\ref{sec:model} by presenting the key features of the vanilla $\lmlt$ model. Section~\ref{sec:emd} is dedicated to explaining the EMD phases in detail, including the evolution of energy density and temperature as functions of the scale factor. We also solve the coupled Boltzmann equations to illustrate DM production in this non-standard freeze-out scenario. DM genesis and direct and indirect detection constraints are presented in Section~\ref{sec:DarkMatter}. In Section~\ref{sec:results}, we present our main findings, followed by a further discussion in Section~\ref{sec:discussions}. Moreover, we summarize our conclusions in Section~\ref{sec:conclusions}. Finally, we reserved Appendix~\ref{app:current} to discuss the current and projected constraints on the leptophilic $Z'$ boson from laboratory experiments, as well as from astrophysical and cosmological observations.

\section{\texorpdfstring{The $\boldsymbol{L_\mu - L_\tau}$ model}{The Lmu - Ltau model}}\label{sec:model}
\begin{figure}[t!]
    \centering
    \includegraphics[width=0.85\linewidth]{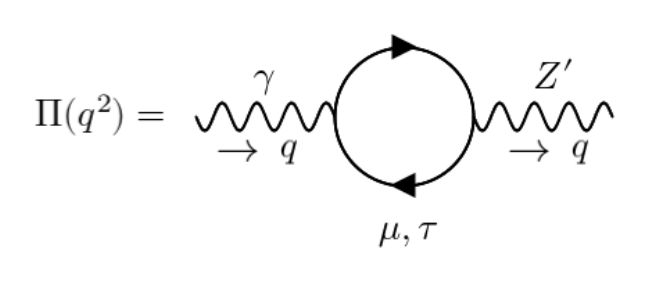}
    \caption{Feynman diagram for the 1-loop kinetic mixing correction.} 
    \label{fig:kineticmixing}
\end{figure}

In this section, we introduce a $U(1)_{\lmlt}$ extension of the SM. In addition to the $Z'$ gauge boson that naturally comes with the new $U(1)_{\lmlt}$ gauge symmetry, this minimal model contains a new SM singlet vector-like Dirac fermion $\chi$, which plays the role of the DM candidate, and a singlet complex scalar $\Phis$ that spontaneously breaks the $\lmlt$ symmetry. The Lagrangian density relevant for the DM phenomenology can be written as~\cite{Altmannshofer:2016jzy, Holst:2021lzm},
\begin{align} \label{eq:vanillaLag}
     \Lag &\supset \bar\chi\left( i \gamma^\mu \partial_\mu - m_\chi \right)\chi + Q_{\mu\tau}^{\chi}\, \gmt\, \bar\chi\, \gamma^\mu\, \chi\, Z'_\mu \nonumber \\
     &\quad + \gmt\, \mathcal{Q}_{\alpha\beta} (\bar{\ell}_\alpha\gamma^\mu \ell_\beta + \bar{\nu}_\alpha \gamma^\mu P_L\nu_\beta)Z'_\mu \nonumber \\
     &\quad - \frac{1}{4} F'^{\mu\nu}F'_{\mu\nu} + \frac{1}{2}m_{Z'} Z'^\mu Z'_\mu - \frac{\varepsilon_0}{2}F^{\mu\nu} F'_{\mu \nu}\,,
\end{align}
where $\mdm$ is the mass of the DM, $\gmt$ is the new gauge coupling, and $Q_{\mu\tau}^{\chi}$ and the matrix $\mathcal{Q}_{\alpha\beta} = {\rm diag}(0,1,-1)$ correspond to the $\lmlt$ charges for the DM and the leptons, respectively~\cite{Chen:2017cic}. Moreover, $\mzp$ stands for the $Z'$ mass, and $F_{\mu \nu}$ and $F^{\prime}_{\mu \nu}$ are the field strengths of the photon and the $Z'$, respectively. Throughout this work, we assume that the tree-level kinetic mixing parameter $\varepsilon_0$ is zero. However, as shown in Fig.~\ref{fig:kineticmixing}, loop contributions cannot be avoided, and at 1-loop they are given by~\cite{Araki:2017wyg, Zhang:2020fiu} 
\begin{align}
    \varepsilon(q^2) &= \varepsilon_0 + \Pi (q^2) \nonumber \\
    &= \frac{8\,e \, \gmt}{(4\pi)^2} \int_0^1 dx\, x(1-x) \ln \frac{\mt^2 - x (1-x)\,q^2}{\mm^2 - x (1-x)\,q^2}\,,
\end{align}
which, in the case where $m_\mu \gg q$, leads to
\begin{equation} \label{eq:kineticmixingvalue}
    \varepsilon \simeq - \frac{e\,\gmt}{12 \pi^2}\, \ln \frac{m_\tau^2}{m_\mu^2} \simeq -  \frac{\gmt}{70} \,.
\end{equation}
This loop correction induces $Z'$ couplings to quarks and electrons and therefore to possible constraints from DM direct detection experiments.

The $Z'$ decays mainly into the second and third families of leptons and DM following~\cite{Wang:2025kit}
\begin{equation}
    \Gamma (Z'\to f\bar{f}) = \frac{k_f\, \gmt^2\, m_{Z'}}{12 \pi} \left(1 + \frac{2\, m_f^2}{m_{Z'}^2}\right) \sqrt{1 - \frac{4\,  m_f^2}{m_{Z'}^2}}
\end{equation}
where $k_f=1$ for $f=\mu$, $\tau$, $\chi$ and $k_f=1/2$ for $f = \nu_\mu$, $\nu_\tau$. However, it can also decay into electrons and quarks, with corresponding partial widths suppressed by the kinetic mixing~\cite{Kaneta:2016uyt, Wang:2025kit}
\begin{align}
    \Gamma(Z'\to e^+e^-) &= \frac{(\varepsilon e)^2 m_{Z'}}{12\pi} \left(1+\frac{2 m_e^2}{m_{Z'}^2}\right)\sqrt{1-\frac{4 m_e^2}{m_{Z'}^2}}, \\
    \Gamma(Z'\to {\rm had}) &=\frac{(\varepsilon e)^2 m_{Z'}}{12\pi} \left(1+\frac{2 m_\mu^2}{m_{Z'}^2}\right)\sqrt{1-\frac{4 m_\mu^2}{m_{Z'}^2}} \nonumber\\
    &\quad\quad \times \,R(s = m_{Z'}^2)\,,
\end{align}
where $R(s)$ is the R-ratio defined by $\sigma(e^+e^- \to {\rm hadrons}) / \sigma(e^+e^- \to \mu^+ \mu^-)$~\cite{ParticleDataGroup:2024cfk}.

In the next sections, we introduce the EMD era and explore the production mechanisms for DM within this cosmological framework. We then present current constraints from direct and indirect detection experiments, focusing on their implications for the $Z'$ boson mass and the $\gmt$ gauge coupling parameter space.

\section{\label{sec:emd}Early matter domination}
Following the end of inflationary reheating, we assume that the universe was composed of SM radiation and a non-relativistic matter component $\phi_m$~\cite{Silva-Malpartida:2023yks}. The evolution of the energy densities of both matter $\rho_m$ and SM $\rho_{\gamma}$ is governed by the set of Boltzmann equations
\begin{align}
    \frac{\mathrm{d} \rho_m}{\mathrm{d} t} + 3\, H\, \rho_m &= - \Gamma_m \, \rho_m\,, \label{boltz_phi} \\
    \frac{\mathrm{d} \rho_{\gamma}}{\mathrm{d} t} + 4\, H\, \rho_{\gamma} &= +\Gamma_m \, \rho_m \,, \label{boltz_R}
\end{align}
where $\Gamma_m$ is the total decay width of $\phi_m$ into SM radiation, $H \equiv \dot{a}/a$ is the Hubble expansion rate and $a$ is the cosmic scale factor. The first Friedmann equation gives
\begin{equation} \label{eq:hub}
    H^2 = \frac{\rho_m + \rho_{\gamma}}{3\, M_P^2}\,,
\end{equation}
with $M_P \simeq 2.4\times 10^{18}$~GeV is the reduced Planck mass. 

We assume that, after cosmic reheating, the early universe was initially dominated by SM radiation. In this scenario, following the timeline of the system, four distinct phases of the evolution of $\rho_{\gamma, m}$ can be identified as follows~\cite{Arias:2019uol, Silva-Malpartida:2024emu}.
\begin{itemize}
    \item[] \textbf{Phase I} ($a<a_{\rm ini}$, $T>T_{\rm ini}$):
    \item[] In this phase, the SM radiation dominates the dynamics of the universe, and since the $\phi_m$ field remains subdominant, the entropy of the SM plasma is conserved.
    \item[] \textbf{Phase II} ($a_{\rm ini}<a<a_{\rm c}$, $T_{\rm ini}>T>T_{\rm c}$):
    \item[] The second phase begins when the energy density of the $\phi_m$ field becomes equal to that of SM-radiation, defining a temperature $T=T_{\rm ini}$ and the associated scale factor $a=a_{\rm ini}$. From then on, the Hubble expansion is driven by $\phi_m$. However, since the decay of $\phi_m$ is still inefficient during this period, the entropy in the SM sector remains effectively conserved and the radiation continues to evolve as a free component, only redshifting.
    \item[] \textbf{Phase III} ($a_{\rm c}<a<a_{\rm fin}$, $T_{\rm c}>T>T_{\rm fin}$):
    \item[] The third phase begins when the $\phi_m$ field starts to efficiently decay into SM particles. At this point, the radiation component no longer evolves as free radiation, as it is continuously sourced from $\phi_m$. This leads to a modification in the scaling of both its energy density and temperature. The system enters a non-adiabatic regime where entropy in the SM plasma is no longer conserved. Despite this, the expansion of the universe remains dominated by $\phi_m$. The phase ends when the Hubble rate becomes comparable to the decay rate of $\phi_m$ that defines a temperature $T_{\rm fin}$. To preserve the success of BBN, this temperature must remain above roughly $T_{\rm BBN} = 4$~MeV~\cite{Kawasaki:2000en, Hannestad:2004px, Cyburt:2015mya, deSalas:2015glj}. Phases II and III together constitute the EMD era; Phase II being adiabatic, while Phase III non-adiabatic.
    \item[] \textbf{Phase IV} ($a_{\rm fin}<a$, $T_{\rm fin}>T$):
    \item[] In the final phase, the universe transitions back to its standard cosmological evolution. This is due to the rapid exponential decay of the $\phi_m$ field, which becomes negligible. From this point on, the Hubble expansion is driven by SM-radiation, and SM-entropy is once again conserved.
\end{itemize}

\begin{figure}[t!]
    \centering
    \includegraphics[width=\linewidth]{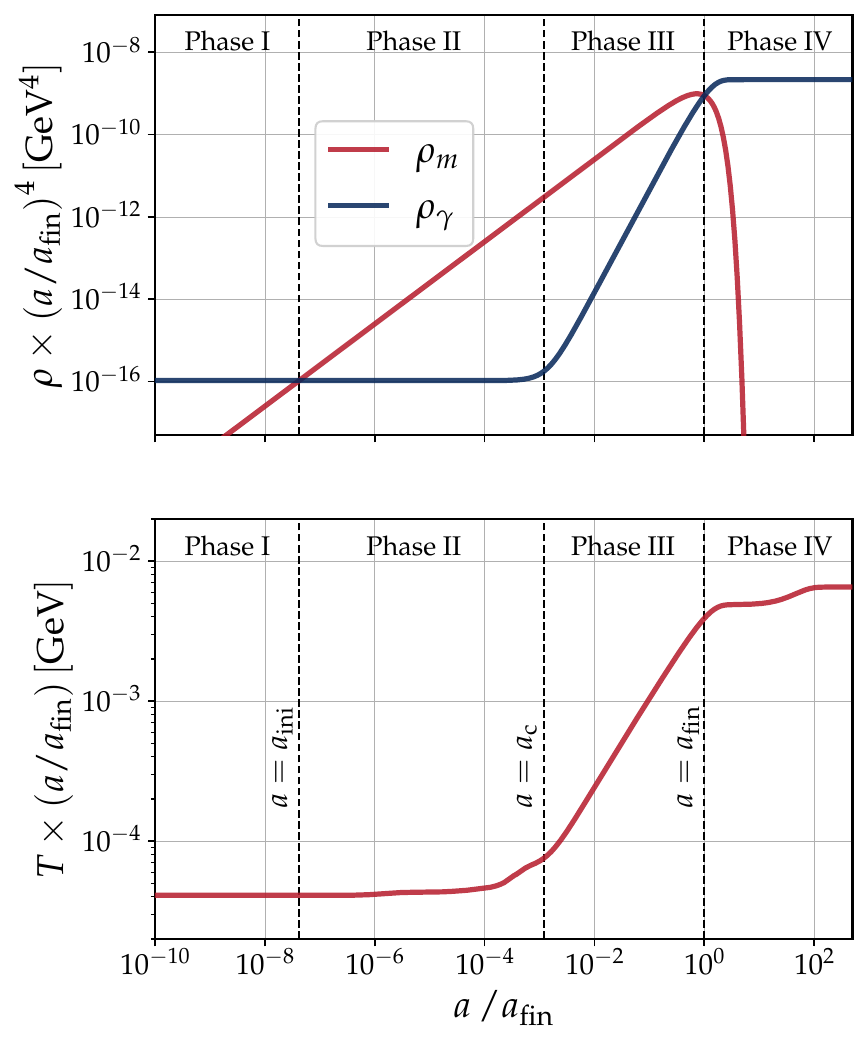}
    \caption{Energy densities (top) and SM-radiation temperature (bottom) as a function of the scale factor $a$, assuming $T_{\rm ini} = 1$ TeV and $T_{\rm fin} = T_{\rm BBN}$.}
    \label{fig:Tend_4mev}
\end{figure}  
We numerically solve Eqs.~\eqref{boltz_phi} and~\eqref{boltz_R}, fixing $T_{\rm fin} = T_{\rm BBN}$ as input parameter to maximize the effect of the EMD era. The results of the numerical solution are presented in Fig.~\ref{fig:Tend_4mev}. In the upper panel, we show the evolution of the energy densities as a function of the scale factor $a$, multiplied by $(a/a_{\rm fin})^4$. This normalization is used because, in the standard cosmological scenario, the SM-radiation energy density scales as $\rho_{\gamma} \propto a^{-4}$. This choice makes it easier to identify deviations from the standard behavior throughout the evolution of the universe. During each phase, the energy densities evolve differently as functions of the $a$, as summarized below:
\begin{equation} \label{eq:scalingR}
    \rho_{\gamma} \propto
    \begin{dcases}
        a^{-4}   & \text{for } a < a_{\rm ini} \hspace{11.1mm}\text{(Phase~I)},\\
        a^{-4} & \text{for } a_{\rm ini} < a < a_{\rm c} \hspace{3.2mm}\text{(Phase~II)},\\
        a^{-3/2} & \text{for } a_{\rm c} < a < a_{\rm fin} \hspace{3mm}\text{(Phase~III)},\\
        a^{-4} & \text{for } a_{\rm fin} < a \hspace{11.3mm}\text{(Phase~IV)},
    \end{dcases}
\end{equation}
while
\begin{equation} \label{eq:scalingm}
    \rho_{m} \propto
    \begin{dcases}
        a^{-3}   & \text{for } a < a_{\rm ini} \hspace{11.1mm}\text{(Phase~I)},\\
        a^{-3} & \text{for } a_{\rm ini} < a < a_{\rm c} \hspace{3.2mm}\text{(Phase~II)},\\
        a^{-3} & \text{for } a_{\rm c} < a < a_{\rm fin} \hspace{3mm}\text{(Phase~III)},\\
        e^{-\Gamma_m/H} & \text{for } a_{\rm fin} < a \hspace{11.3mm}\text{(Phase~IV)}.
    \end{dcases}
\end{equation}
The SM radiation energy density is expressed as a function of the SM temperature as
\begin{equation}
    \rho_{\gamma} (T) = \frac{\pi^2}{30}\, g_\star(T)\, T^4,
\end{equation}
where $g_\star$ is the effective number of relativistic degrees of freedom contributing to the energy density.\footnote{An important observation is that the QCD transition in the effective degrees of freedom of the radiation energy density occurs at $T \sim 100$ MeV, as can be seen in the lower panel of Fig.~\ref{fig:Tend_4mev}} In the lower panel of Fig.~\ref{fig:Tend_4mev}, we show the evolution of the SM radiation temperature $T$, normalized by $a/a_{\rm fin}$, as a function of the scale factor. This allows us to clearly observe how the temperature evolves during each phase of the expansion following distinct scalings
\begin{equation} \label{eq:scalingT}
    T \propto
    \begin{dcases}
        a^{-1}   & \text{for } a < a_{\rm ini} \hspace{11.1mm}\text{(Phase~I)},\\
        a^{-1} & \text{for } a_{\rm ini} < a < a_{\rm c} \hspace{3.2mm}\text{(Phase~II)},\\
        a^{-3/8} & \text{for } a_{\rm c} < a < a_{\rm fin} \hspace{3mm}\text{(Phase~III)},\\
        a^{-1} & \text{for } a_{\rm fin} < a \hspace{11.3mm}\text{(Phase~IV)}.
    \end{dcases}
\end{equation}

The distinct stages of this cosmological scenario significantly alter the thermal evolution of the universe, modifying the characteristics of DM freeze-out. These changes affect the resulting relic abundance and thus the parameter space required for DM to account for current observations. We explore these effects in the next section.

\section{Dark Matter} \label{sec:DarkMatter}
In this section, the DM production in the early universe and the detection prospects will be carefully studied.

\subsection{Dark matter genesis}
In this work, we consider a thermal DM candidate whose cosmological abundance is determined by the freeze-out mechanism. Accordingly, the evolution of its total number density, defined as $n\equiv n_{\chi}+n_{\bar{\chi}}$, is governed by the Boltzmann equation~\cite{Holst:2021lzm}\footnote{We neglect the possibility of direct decays of the field $\phi_m$ into $\chi$, which is a valid approximation provided that the branching ratio Br$_{\phi_m \to \chi \bar{\chi}}$ remains below approximately $10^{-4}~m_{\chi}/(100~\text{GeV})$~\cite{Drees:2017iod, Arias:2019uol}.}
\begin{equation} \label{eq:boltzdm}
    \frac{dn}{dt} + 3\, H\, n = - \frac{\langle\sigma v\rangle}{2} \left(n^2 - n_\text{eq}^2\right),
\end{equation}
where $n_{\text{eq}}$ corresponds to the equilibrium DM number density and $\langle\sigma v\rangle$ is the 2-to-2 thermally-averaged DM annihilation cross-section described by~\cite{Gondolo:1990dk}
\begin{equation} \label{eq:sv}
     \langle\sigma v\rangle = \int_{4m_{\chi}^2}^{\infty} \mathrm{d}s\, \frac{(s-4m_{\chi}^2) \, \sqrt{s} \, K_{1}\left ( \sqrt{s}/T \right ) \sigma(s)}{8 \,T\, m_{\chi}^{4}\, K_2^2\left ( m_{\chi}/T \right ) }\,,
\end{equation}
where $K_{i}$ denotes the modified Bessel function, and $\sigma$ represents the 2-to-2 cross-section. DM can annihilate into fermions $\left ( \chi \bar{\chi} \to f \bar{f} \right)$ through the $s$-channel exchange of a $Z'$ or into a pair of $Z'$ $\left ( \chi \bar{\chi} \to Z' Z' \right )$ via the $t$-channel exchange of a DM particle, if these processes are allowed kinematically.

We generate all relevant annihilations and compute the total cross-section using the \texttt{standalone} subroutine from \texttt{MadGraph}~\cite{Alwall:2014hca}, where we include the \texttt{UFO} file generated by \texttt{FeynRules}~\cite{Alloul:2013bka} for our model. Then, we perform the numerical integration of $\langle\sigma v\rangle$ (cf. Eq.~\eqref{eq:sv}) using an \texttt{implicit Euler} technique. Finally, we numerically solve the Boltzmann equation for DM (that is, Eq.~\eqref{eq:boltzdm}) for a specific \texttt{param\_card}, using adaptive numerical methods.

\begin{figure}[t!]
    \centering
    \includegraphics[width=\linewidth]{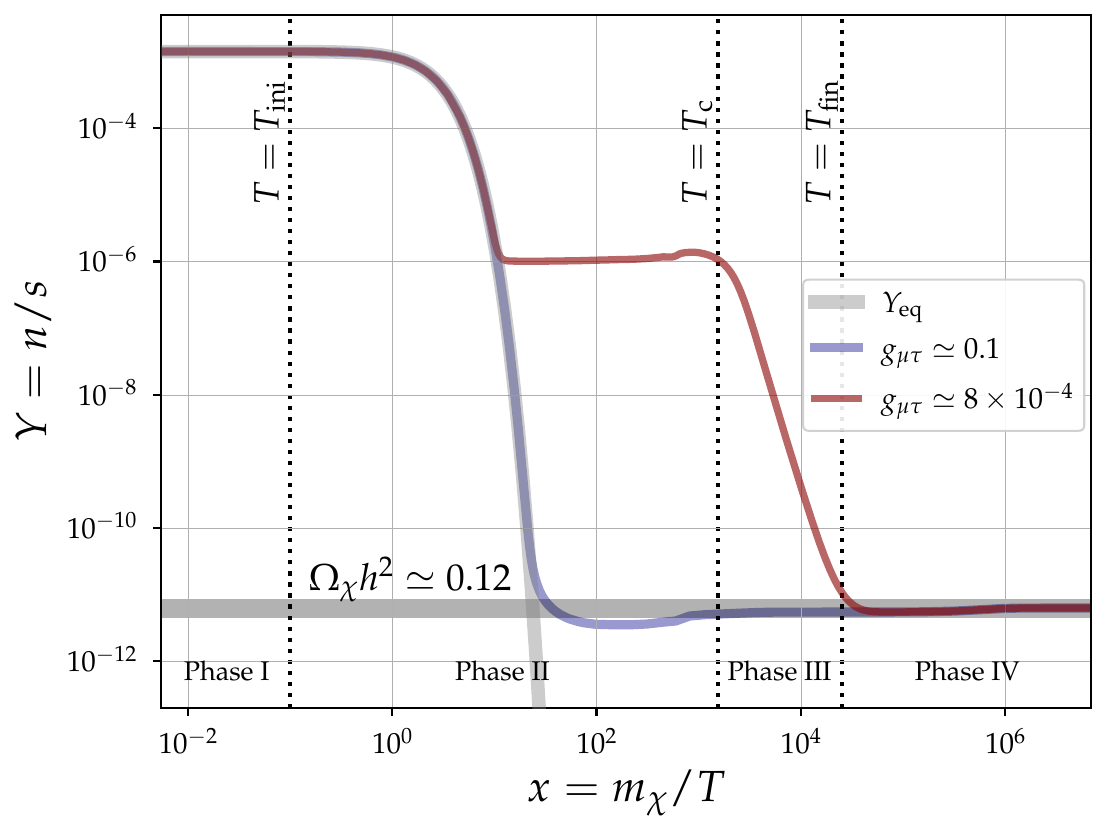}
    \caption{DM yield as a function of the inverse SM-radiation temperature. The thick horizontal line represents the observed DM relic abundance. The red curve shows the DM evolution including the effect of EMD, while the blue curve corresponds to the standard cosmological scenario. In this plot, we set $m_\chi = 100$~GeV, $m_{\chi} / m_{Z'} = 0.3$, $T_{\rm ini} = 1$~TeV and $T_{\rm fin} = T_{\rm BBN}$.}
    \label{fig:yield}
\end{figure}  
An example of the effect of an EDM era on the genesis of DM is illustrated in Fig.~\ref{fig:yield}, assuming $T_{\rm ini}=1$~TeV and $T_{\rm fin} = T_{\rm BBN}$ (which gives a critical temperature $T_{\rm c} \simeq 60$~MeV). We also took a DM mass $m_{\chi} = 100$~GeV, $m_\chi / m_{Z'} = 0.3$ and a gauge coupling $\gmt \simeq 5 \times 10^{-4}$ to fit the whole observed DM abundance. The figure shows the evolution of the yield $Y\equiv n/s$, where $s$ is the entropy density of the SM-radiation plasma, as a function of the inverse temperature variable $x\equiv m_{\chi}/T$. In this scenario, the DM (solid red line) decouples from the thermal equilibrium bath $Y_{\rm eq}$ (solid gray line) towards the end of Phase II. The decay of the matter field $\phi_m$ into SM-radiation, occurring between $T_{\rm c}$ and $T_{\rm fin}$, leads to a significant injection of entropy into the SM plasma. As a result, the DM yield is substantially diluted during Phase III. To reproduce the observed relic abundance ($\Omega_{\chi}\, h^2 \simeq 0.12$~\cite{Planck:2018vyg}), the yield must be enhanced, which requires reducing the annihilation cross section $\langle\sigma v\rangle$. This, in turn, implies a smaller coupling $\gmt$, compared to the standard scenario (red dashed line). The coupling in the scenario affected by EMD ($\gmt \simeq 8 \times 10^{-4}$) is about two orders of magnitude smaller than the one in the standard case ($\gmt \simeq 0.1$).

\subsection{Direct detection} \label{secc:direct}
\begin{figure}[t!]
    \centering
    \includegraphics[width=\linewidth]{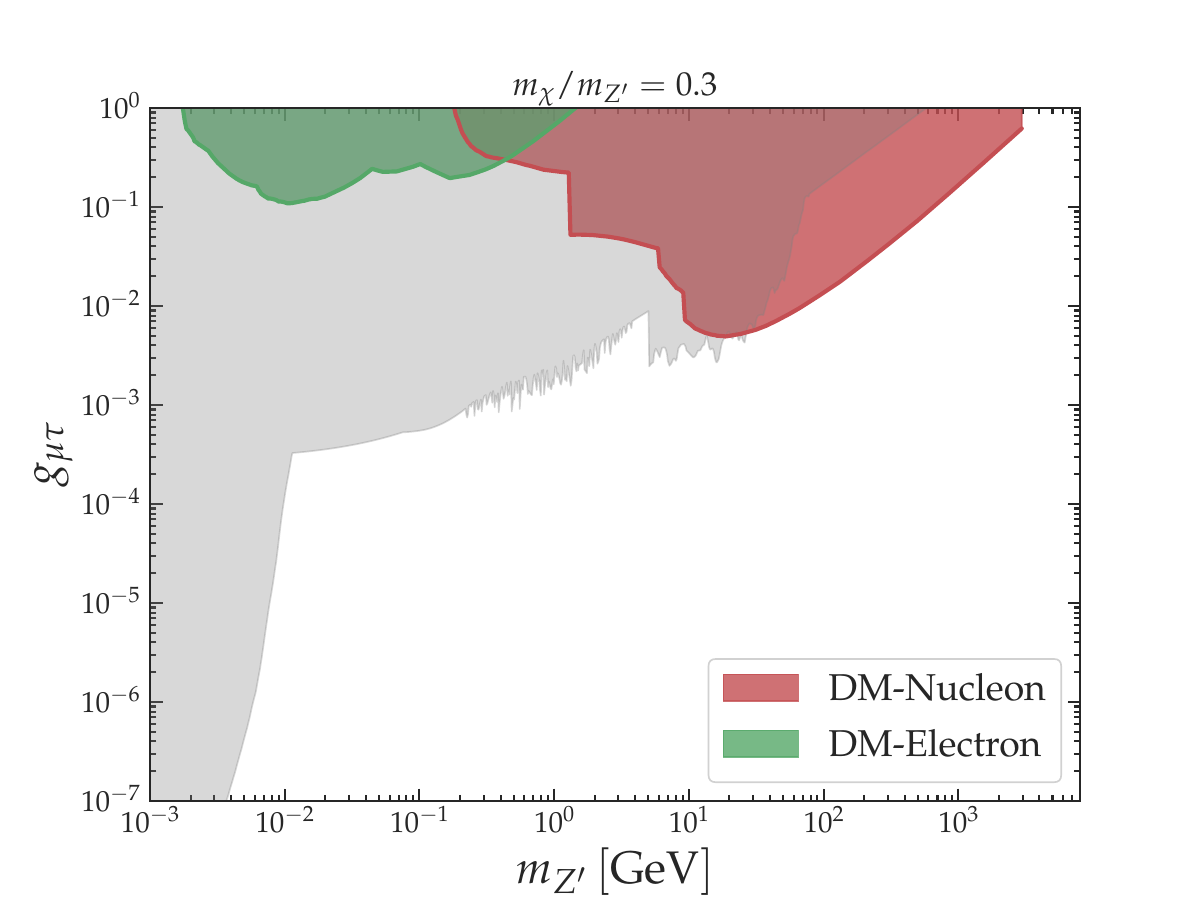}
    \includegraphics[width=\linewidth]{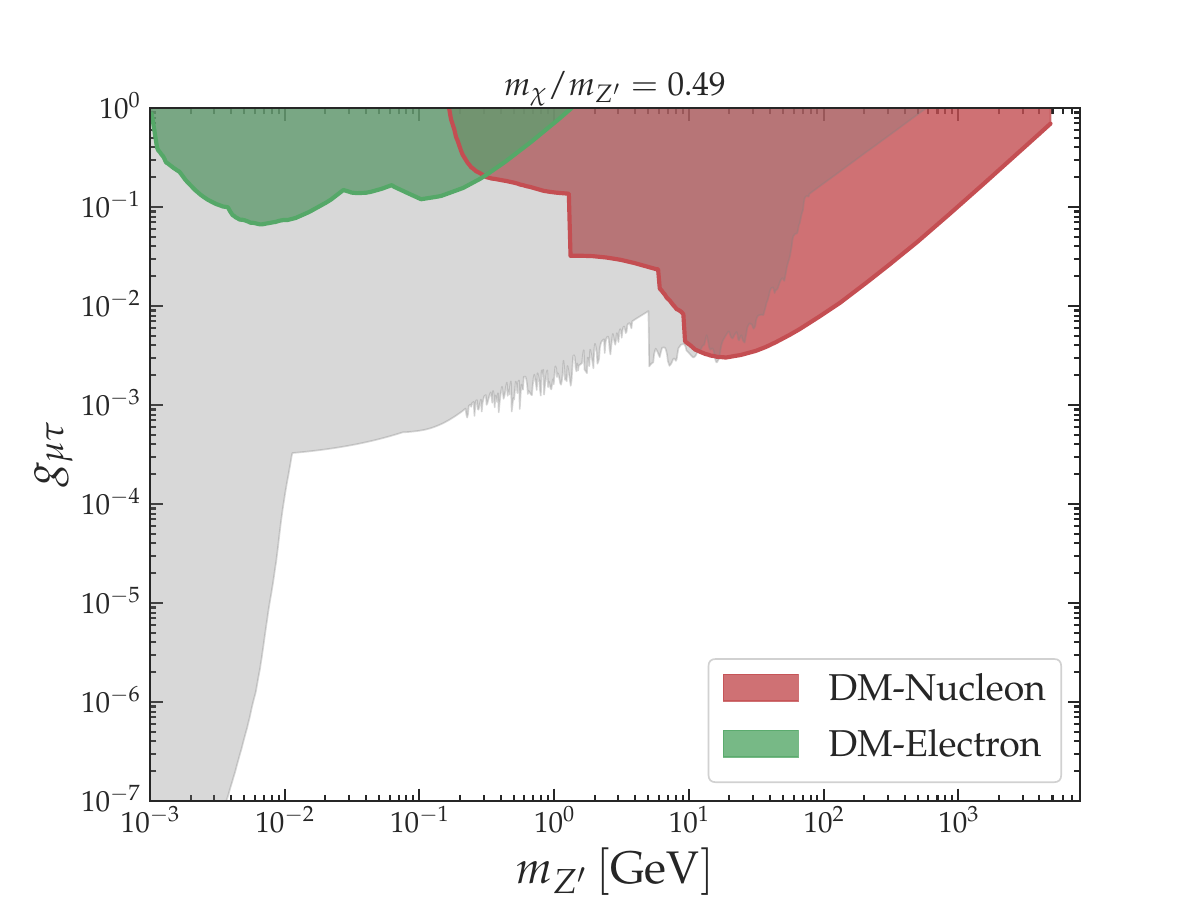}
    \caption{Bounds from DM direct detection stemming from DM-nucleon scattering (red) and DM-electron scattering (green), for different values of $m_\chi/m_{Z'}$. In gray, we show the $Z'$ boson experimental and observational bounds, see Appendix~\ref{app:current}.}
    \label{fig:dd_zp_bounds}
\end{figure}

The DM-nucleon and DM-electron interactions arise at loop level due to the loop-induced kinetic mixing, since the $Z'$ boson does not interact directly with electrons and quarks. The spin-independent (SI) DM-nucleon scattering cross-section is given by
\begin{equation}
    \sigma_{\chi N}^{\rm SI} = \frac{\mu_{\chi N}^2}{\pi} \left(\frac{Z}{A}\right)^2 \frac{\gmt^2 \epsilon^2 e^2}{m_{Z'}^4}\,,
\end{equation}
where the DM-nucleon reduced mass is given by $\mu_{\chi N} \equiv m_\chi\, m_N / (m_\chi + m_N)$, and $A$ and $Z$ are the mass and atomic number of the nucleus, respectively. Additionally, the SI DM-electron scattering cross-section is given by~\cite{Figueroa:2024tmn, Wang:2025kit}
\begin{equation}
    \sigma_{\chi e} = \frac{\mu_{\chi e}^2}{\pi}\, \frac{\gmt^2 \epsilon^2 e^2}{(m_{Z'}^2 + \alpha^2m_e^2)^2}\,,
\end{equation}
where $\mu_{\chi e} \equiv m_\chi m_e / (m_\chi + m_e)$ is the reduced mass for the DM-electron system and $\alpha$ is the fine structure constant. Figure~\ref{fig:dd_zp_bounds} presents in red the combined bounds from different DM-nucleon direct detection experiments: CRESST-III~\cite{CRESST:2019jnq}, DarkSide-50~\cite{DarkSide-50:2022qzh}, LUX-ZEPLIN~\cite{LZ:2024zvo}, and XENONnT~\cite{XENON:2023cxc}, in the plane $[m_{Z'},\, \gmt]$, for $m_\chi / m_{Z'} = 0.3$ (top) and $m_\chi / m_{Z'} = 0.49$ (bottom). In addition, the bounds from DM-electron scattering are shown in green and come from SENSEI~\cite{SENSEI:2020dpa, SENSEI:2024yyt}, PANDAX-II~\cite{PandaX-II:2021nsg}, and XENON1T~\cite{XENON:2019gfn}. The bounds provided by the DM-electron scattering are not competitive and correspond to a parameter space that has already been constrained by other experiments. However, in the GeV mass range, the DM-nucleon scattering excludes a relevant region for $\gmt \gtrsim 10^{-3}$.

Finally, the gray area corresponds to the combination of other constraints on the model, beyond DM. It contains bounds from the effective number of neutrinos, cooling of white dwarfs, the muon anomalous magnetic moment, borexino, neutrino trident production, ATLAS, NA64$\mu$, BaBar and Belle-II. Details on the derivation of the limits are reported in the Appendix~\ref{app:current}.

\subsection{Indirect detection} \label{secc:indirect}
DM annihilations into SM particles can be indirectly probed through high-energy gamma-ray observations and cosmic microwave background (CMB) spectral measurements.

\begin{figure}[t!]
    \centering
    \includegraphics[width=\linewidth]{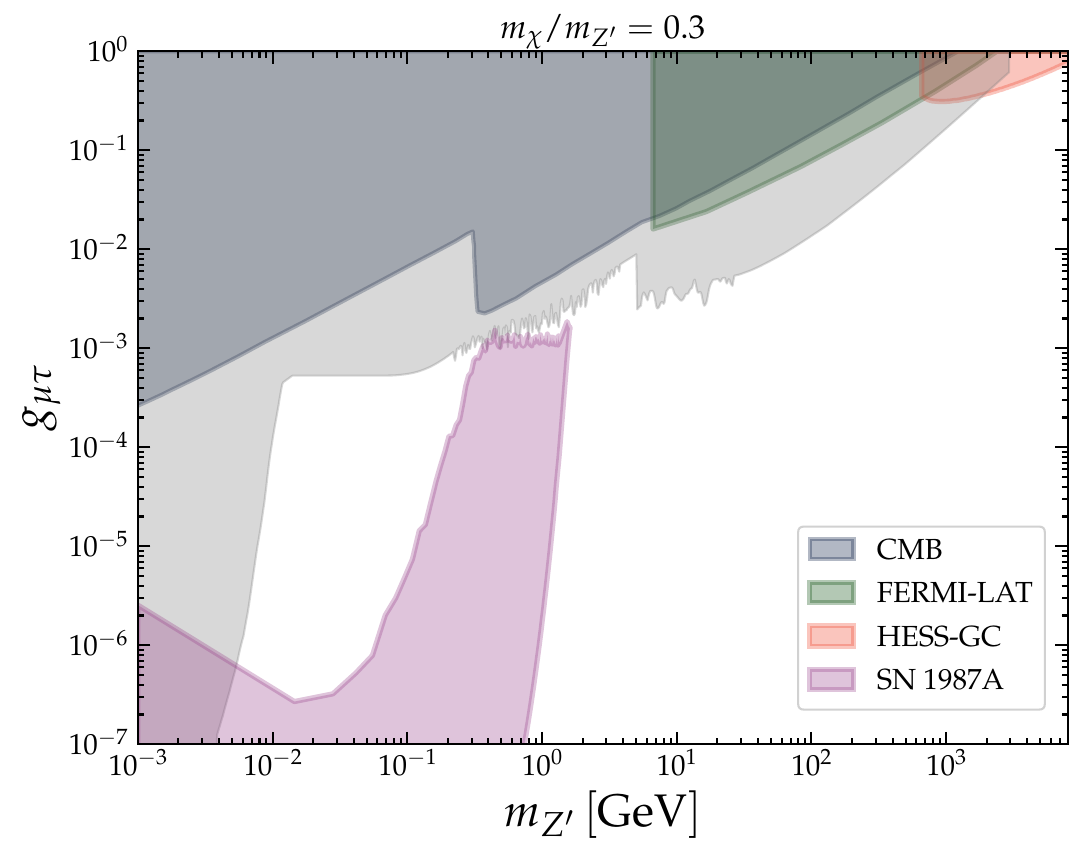}
    \includegraphics[width=\linewidth]{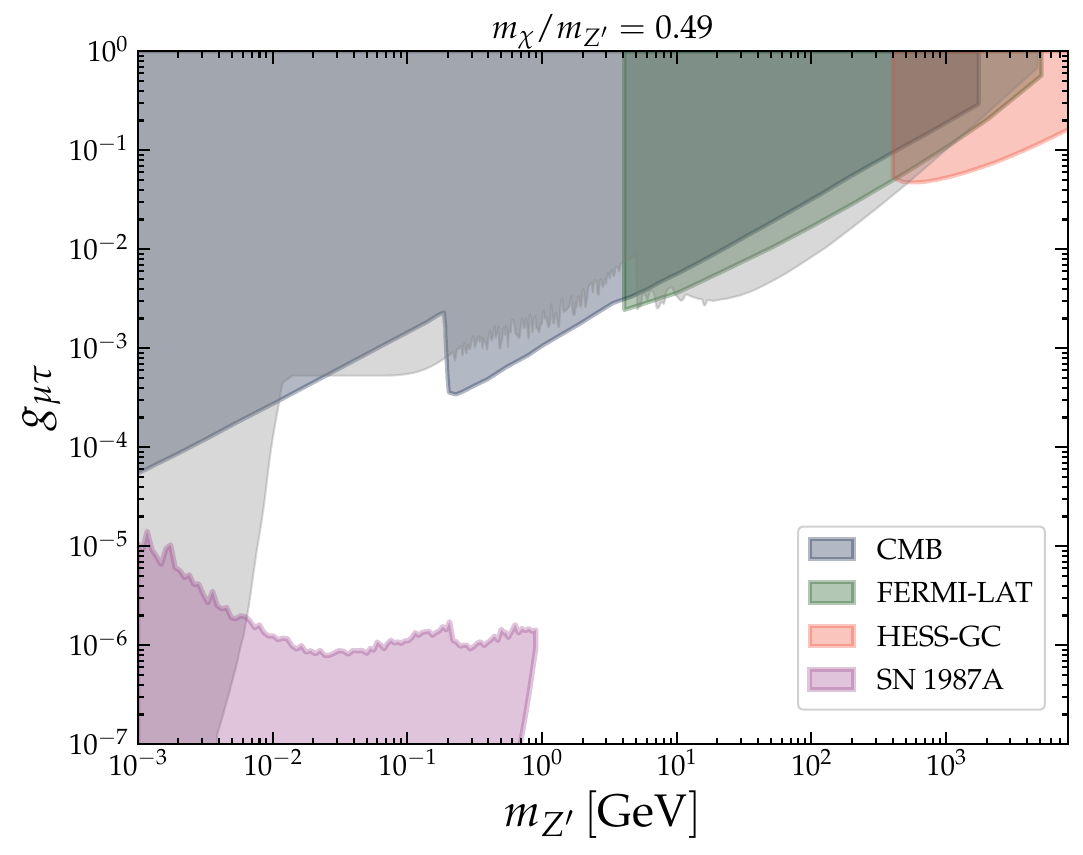}
    \caption{Bounds from DM indirect detection, from Fermi-LAT (green), H.E.S.S. (red), CMB anisotropies (blue), and Supernova cooling (magenta). Lastly, the gray region shows the $Z'$ boson experimental and observational bounds, see Appendix~\ref{app:current}, combined with the DM direct detection limits.}
    \label{fig:id_zp_bounds}
\end{figure}

On the gamma-ray side, detectors such as Fermi-LAT and H.E.S.S. are sensitive to the photon spectrum resulting from DM annihilations into charged SM fermions. In our model, such annihilation channels predominantly produce charged leptons, which subsequently lead to continuum gamma-ray emission. The total differential gamma-ray flux with respect to the energy $E_\gamma$ of the photon and the solid angle $\Delta \Omega$ is the sum of the contributions from individual lepton channels and can be written as
\begin{equation}
    \frac{d \Phi_\gamma}{dE_\gamma \, d\Omega} = \frac{\langle \sigma v \rangle}{8 \pi m_\chi^2} \frac{dN_\gamma}{dE_\gamma} \, J_{\rm ann}(\Delta\Omega)\, ,
\end{equation}
where the $J$-factor, $J_{\rm ann}(\Delta\Omega)$, encodes the DM distribution along the line of sight and depends quadratically on the DM density profile. The photon energy spectrum per annihilation is given by
\begin{equation}
    \frac{dN_\gamma}{dE_\gamma} = \sum_{\ell = \mu, \tau} {\rm BR}_\ell \frac{dN_\gamma^{\ell}}{dE_\gamma}\,,
\end{equation}
where ${\rm BR}_\ell$ denotes the fraction of the annihilation channel $\chi\chi \to \ell^+ \ell^-$ relative to the total annihilation cross-section $\langle \sigma v \rangle_{\rm tot}$. We neglect the primary $e^+e^-$ final state, as its contribution is suppressed due to the 1-loop kinetic mixing to electrons. We use the limits on the annihilation cross-section as a function of the DM mass released by the collaborations to constrain the parameter space of our model. H.E.S.S. used the Einasto profile for the galactic center with a local DM density $\rho_0 = 0.4$~GeV/cm$^3$, while Fermi-LAT assumes a Navarro-Frenk-White profile for dwarf spheroidal galaxies. The resulting gamma-ray constraints are shown as green and red shaded regions in Fig.~\ref{fig:id_zp_bounds}, corresponding to Fermi-LAT based on the analysis of gamma rays from dwarf spheroidal galaxies~\cite{Fermi-LAT:2015att}, and H.E.S.S. from searches for DM annihilation in the galactic center~\cite{HESS:2022ygk}, respectively. Within the GeV-scale DM mass range, Fermi-LAT provides the most stringent limits from the $\tau^+ \tau^-$ channel, up to about $100$~GeV. For higher masses, H.E.S.S. becomes the dominant probe, especially for annihilation into tau leptons.

In addition, the CMB provides a complementary constraint: DM annihilations into charged particles can increase the ionization fraction, injecting energy into the plasma during the recombination era and thus modifying the CMB anisotropy spectrum~\cite{Leane:2018kjk}. The relevant quantity governing this effect is
\begin{equation}
    p_{\rm ann} = \sum_{\ell = e,\mu,\tau} \frac{1}{2}f_{\rm eff}^{\ell} \, {\rm BR}_\ell \frac{\langle \sigma v \rangle}{m_\chi}\,,
\end{equation}
where $f_{\rm eff}^{\ell}$ is the efficiency factor for the deposition of energy by each lepton flavor $\ell$~\cite{Slatyer:2015jla}.
Based on this parameterization, the Planck collaboration has established the upper bound~\cite{Planck:2018vyg}
\begin{equation}
    p_{\rm ann}  < 3.5 \times 10^{-28} {\rm cm}^{3}\,{\rm s}^{-1}\, {\rm GeV}^{-1}.    
\end{equation}
This constraint is shown as the blue region in Fig.~\ref{fig:id_zp_bounds}. The suppressed annihilation into $e^+e^-$ pairs gives rise to a change in the slope of the CMB exclusion curve for $m_\chi < m_\mu$.

From astrophysics, we can also account for the limits from supernova cooling. It is expected that a thermal population of muons exists within the dense and hot environment of proto-neutron stars, due to the presence of trapped neutrinos. Within the $\lmlt$ model, DM particles can be produced through the new leptophilic $Z'$ portal. If DM escapes from the proto-neutron stars, leading to a significant loss of energy, its dark luminosity $L_\chi$ could be constrained by comparing it with the neutrino luminosity $L_\nu$, 
\begin{equation}
    L_\chi \leq L_\nu = 5.7 \times 10^{52} \,\,{\rm erg}\,\,{\rm s}^{-1}\,.
\end{equation}
In Ref.~\cite{Manzari:2023gkt}, the authors derived stringent constraints over the general $\lmlt$ parameter space from the observation of a neutrino signal in the supernova SN 1987A in the Large Magellanic Cloud (LMC)~\cite{Kamiokande-II:1987idp, Bionta:1987qt, Alekseev:1988gp, Loredo:2001rx, Vissani:2014doa}, using the supernovae simulation data of the radial profile labeled as SFHo-18.8 and SFHo-20.0~\cite{Woosley:2007as, Sukhbold:2017cnt}.

We used the code provided by Ref.~\cite{Manzari:2023gkt} to obtain the limits shown in magenta in Fig.~\ref{fig:id_zp_bounds} for the $\lmlt$ model, according to the mass ratios $m_\chi/m_{Z'}$ of our interest.\footnote{The code is available on \texttt{\href{https://github.com/spinjo/SNforMuTau}{SNforMuTau.git}}.} Notably, these supernova cooling bounds can probe complementary regions of the parameter space, especially for the sub-GeV DM mass range.

\begin{figure*}[t!]
    \centering
    \includegraphics[width=.49\linewidth]{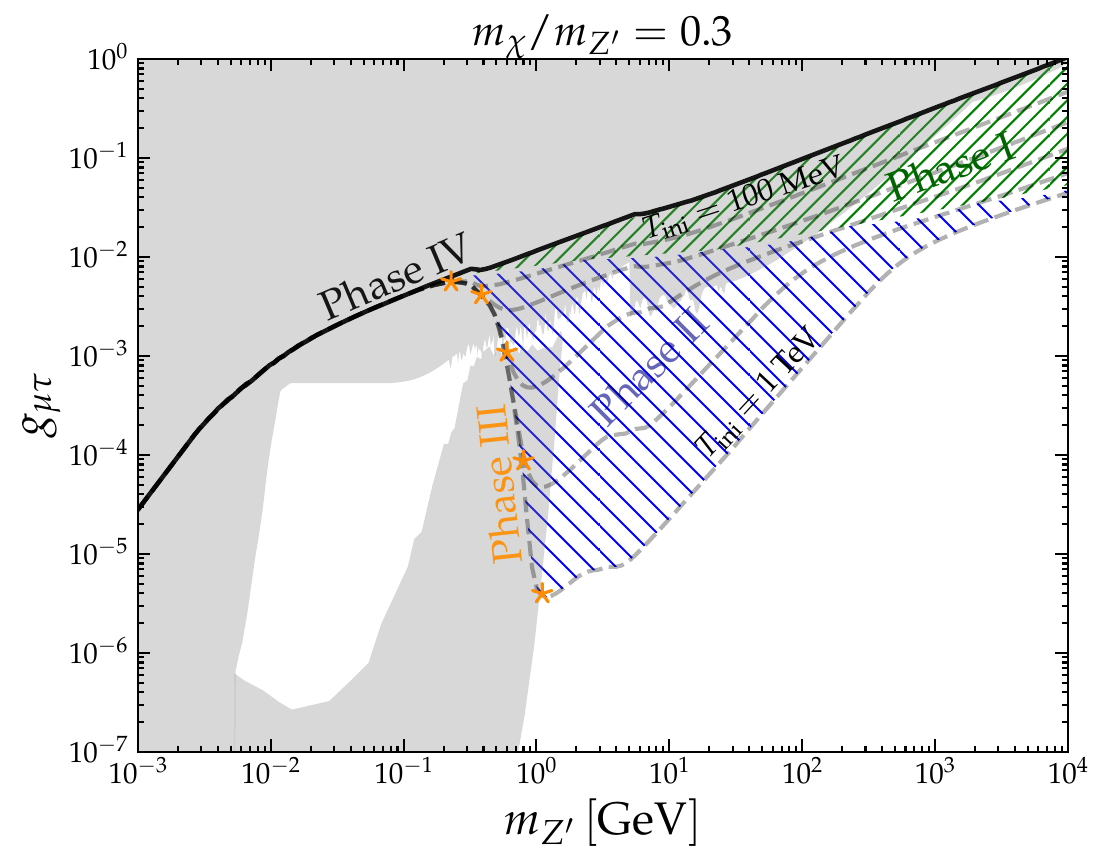}
    \includegraphics[width=.49\linewidth]{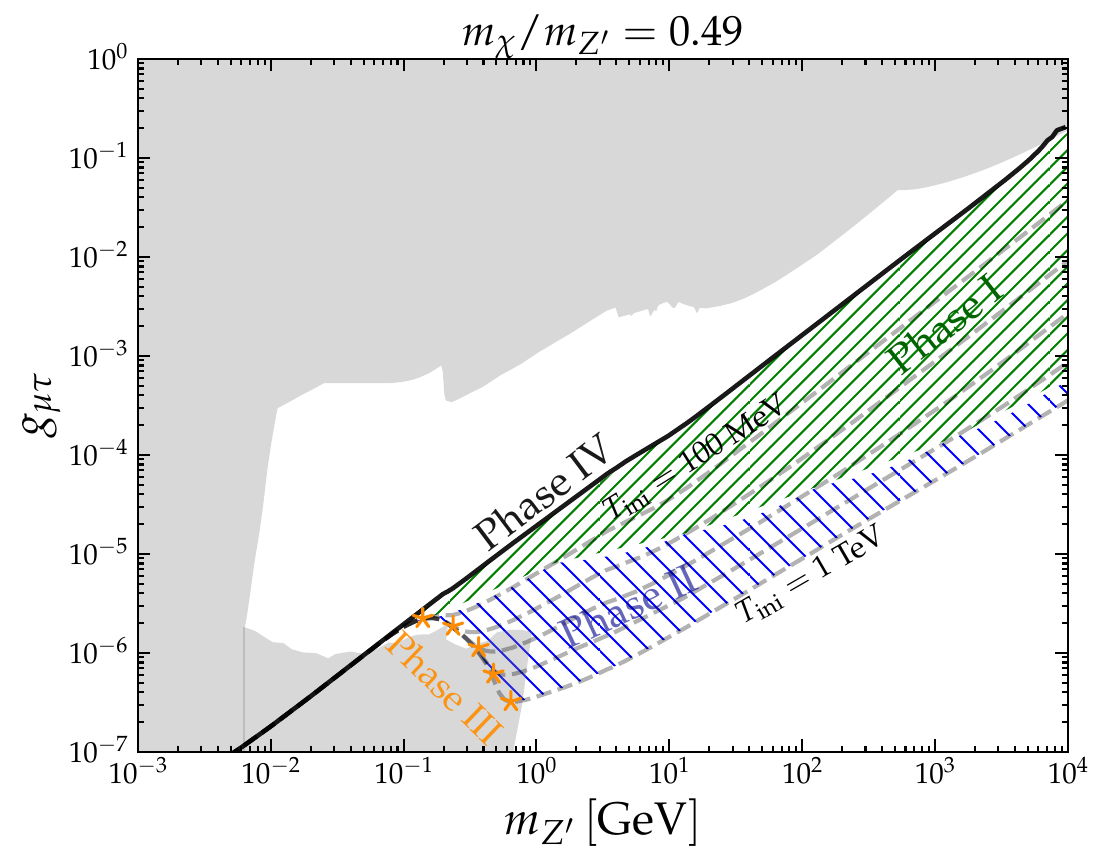}
    \caption{Parameter space that fits the whole observed DM abundance in an EMD, for $T_{\rm fin} = T_{\rm BBN}$. The left (right) panel corresponds to $m_\chi/m_Z = 0.3$ ($0.49$). The green and blue regions indicate the portions of parameter space where the effects of Phases I and II are relevant, respectively. Phase III lies upon the line connecting the \textcolor{orange}{$\star$} symbols. The solid black line represents Phase IV, where the observed DM relic density is achieved, and freeze-out occurs entirely after the EMD era has ended; i.e., under standard cosmology with no non-standard effects. Each gray dashed line corresponds to the correct DM abundance assuming an EMD era with different initial temperatures: $T_{\rm ini} = 100$~MeV, 1~GeV, 10~GeV, 100~GeV, and 1~TeV.}
    \label{fig:relic_rate_0.3}
\end{figure*}
\section{Dark Matter in an EMD era} \label{sec:results}
In this section, we examine the impact of an EMD era on the parameter space that fits the observed DM relic abundance in the context of the $L_\mu - L_\tau$ model, within the thermal freeze-out framework.

Figure~\ref{fig:relic_rate_0.3} shows the results in the $[m_{Z'},\, g_{\mu\tau}]$ plane for a fixed $T_{\rm fin} = T_{\rm BBN}$. The left (right) panel corresponds to $m_\chi/m_{Z'} = 0.3$ ($0.49$), with the right panel lying close to the resonant regime. The thick black solid and blue dashed lines correspond to the parameter values that reproduce the observed relic abundance $\Omega_\chi h^2 \simeq 0.12$. The solid line refers to the standard scenario in which DM decouples after the end of the EMD era, that is, during the radiation-dominated phase (denoted Phase IV). In this case, the freeze-out occurs at $T_{\rm fo} < T_{\rm fin}$. The dashed lines, on the other hand, represent scenarios in which DM decouples during or before the EMD epoch, that is, at $T_{\rm fo} > T_{\rm fin}$. We consider several onset temperatures for the EMD phase: $T_{\rm ini} = 100$~MeV, 1~GeV, 10~GeV, 100~GeV, and 1~TeV. The gray-shaded regions indicate the current experimental constraints on the model and contain direct and indirect DM detection limits, together with the non-DM bounds in the Appendix~\ref{app:current}.

We begin by discussing the different cosmological phases that appear across both panels of Fig.~\ref{fig:relic_rate_0.3}, focusing on how they shape the relic density. The green-hashed region corresponds to Phase~I, where DM freezes out during the initial radiation-dominated era, at temperatures $T_{\rm fo} > T_{\rm ini}$. This phase mimics standard freeze-out behavior in radiation domination, but requires smaller couplings as a result of the {\it maximal} entropy dilution from the subsequent EMD period. This effect manifests itself as curves with the same slope as in the standard case within the green region.

The blue-hashed region corresponds to Phase II, where the DM decouples during the adiabatic EMD epoch that has already begun to alter the cosmic expansion rate ($T_{\rm ini} > T_{\rm fo} > T_{\rm c}$), but injection of entropy into the SM plasma has not yet started. The change in the slope of the relic density contours here is due to the modified Hubble rate (see Eq.~\eqref{eq:hub}).

Phase III is marked by the orange star-shaped trajectory (\textcolor{orange}{$\star$}) and corresponds to freeze-out occurring while entropy is actively being injected into the SM plasma, i.e., $T_{\rm c} > T_{\rm fo} > T_{\rm fin}$. In this regime, the plasma temperature evolves more slowly, scaling as $T \propto a^{-3/8}$ rather than the usual $a^{-1}$, significantly affecting the annihilation rate required to obtain the correct relic density. This effect becomes more pronounced as $T_{\rm ini}$ increases, resulting in steeper dashed lines. The earlier the onset of EMD, the larger its impact on the expansion rate, which in turn influences the coupling $g_{\mu\tau}$ needed for efficient DM annihilation.

\begin{figure*}[t!]
    \centering
    \includegraphics[width=0.49\linewidth]{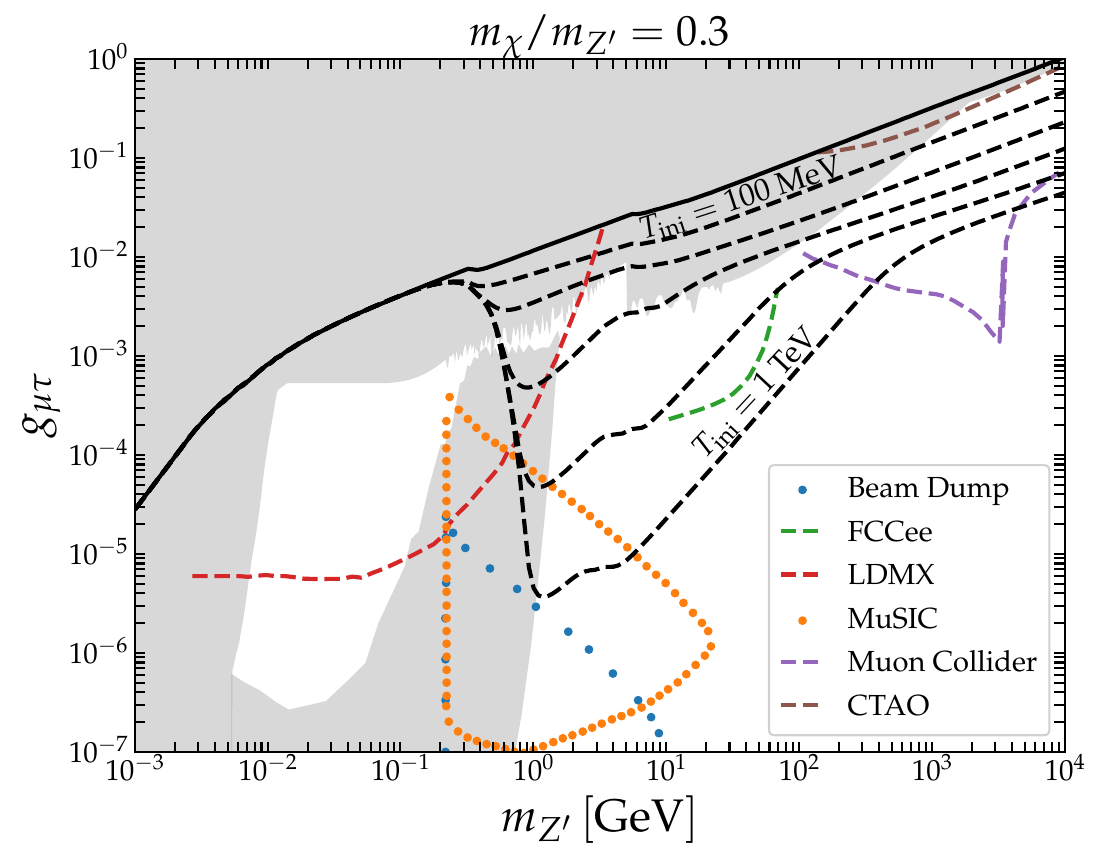}
    \includegraphics[width=0.49\linewidth]{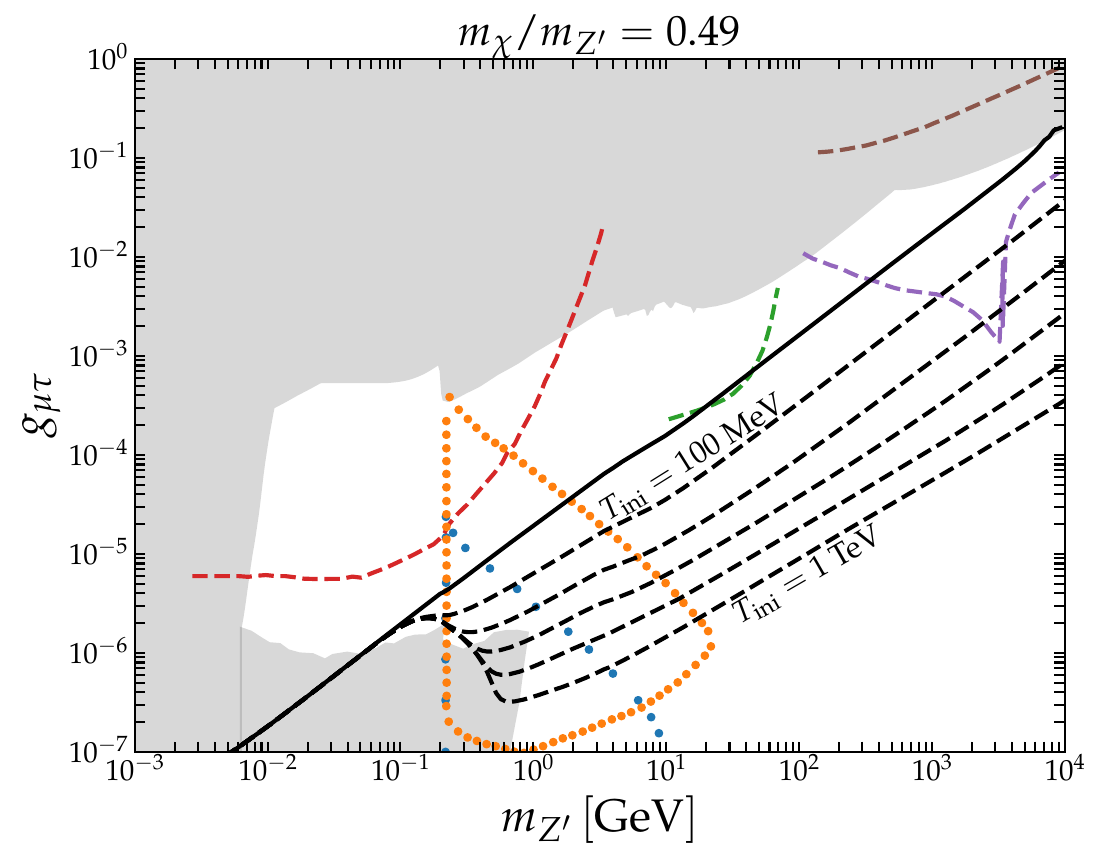}
    \caption{Similar to Fig.~\ref{fig:relic_rate_0.3}, but overlaying projected sensitivities of future detectors with colorful dots and dashed curves.}
    \label{fig:relic_rate_0.49}
\end{figure*}
It is worth highlighting that across Phases I to III, the coupling $g_{\mu\tau}$ required to fit the observed DM relic abundance is reduced compared to the case dominated by SM radiation. This suppression comes from the injection of entropy that dilutes the DM number density. Interestingly, as smaller couplings are explored, in some cases it is possible to escape the strongest experimental constraints shown in gray. For example, in the case with $m_\chi/m_{Z'} = 0.3$, the standard cosmological history is in tension with existing constraints. However, the inclusion of an EMD era revitalizes this mass ratio, rendering it completely viable.

Figure~\ref{fig:relic_rate_0.49} resembles Fig.~\ref{fig:relic_rate_0.3} as it shows the parameter space that fits the total observed DM abundance with solid and dashed black lines, and the areas already excluded by the present measurements (gray areas). However, it also overlays projected sensitivities of future detectors. The blue dots indicate the expected sensitivity of beam-dump experiments with accelerated muon beams~\cite{Cesarotti:2023sje}. The green dashed curve corresponds to the sensitivity of the FCC-ee~\cite{Airen:2024iiy}, while the red dashed curve represents the projected reach of LDMX~\cite{Kahn:2018cqs}. In orange dots, we present the MuSIC prospects~\cite{Davoudiasl:2024fiz}, while the magenta dashed line shows the expected sensitivity of the muon collider~\cite{Huang:2021nkl}. Finally, the brown dashed line corresponds to the CTAO sensitivity~\cite{CTA:2020qlo}, and the pink dashed curve displays the projected reach of DARWIN~\cite{DARWIN:2016hyl, Baudis:2024jnk}. The details of the derivation of such bounds are presented in the Appendix~\ref{app:current}. It is interesting to note that EMD allow escape of current stringent experimental bounds but favor regions of the parameter space that could be probed with next-generation experiments.

There is no doubt that the vanilla $U(1)_{\lmlt}$ model offers a compelling framework for investigating the origin of DM. Beyond its applications in standard cosmology, this model also provides an avenue to probe non-standard cosmological histories. In the next section, we further discuss the implications of non-standard cosmological scenarios.

\section{Discussions} \label{sec:discussions}
In the standard thermal freeze-out scenario, certain parameter choices, such as a mass ratio of $m_\chi/m_{Z'} = 0.3$, are completely excluded by current experimental constraints (cf. the left panel of Fig.~\ref{fig:relic_rate_0.3}). However, a modification of this ratio allows one to evade the bounds. For example, on the right panel of Fig.~\ref{fig:relic_rate_0.3}, we present the results for $m_\chi/m_{Z'} = 0.49$. This choice not only circumvents existing limits but also yields predictions that are within the reach of future experimental sensitivities (cf. the right panel of Fig.~\ref{fig:relic_rate_0.49}). Notice that as we increase the $m_\chi/m_{Z'}$ ratio closer to 0.5 the model relies on the $Z'$ resonance, allowing us to have smaller gauge couplings and consequently evading current bounds. At this point, one might naturally ask: Why consider non-standard cosmological scenarios? Let us elaborate.

Below the solid black curve, the predicted DM relic abundance exceeds the observed value $\Omega_\chi h^2 > 0.12$, overclosing the universe. As the ratio $m_\chi/m_{Z'}$ approaches the resonance region, the relic density saturates, requiring increasingly smaller gauge couplings to match observations. For example, with $m_\chi/m_{Z'} = 0.49$, couplings as small as $10^{-6}$ or $10^{-7}$ and $Z'$ masses near 5~GeV fall outside the reach of standard freeze-out, effectively ruling out this parameter space in conventional cosmology. In contrast, these scenarios arise naturally in the presence of an EMD phase, which restores their phenomenological viability. The EMD epoch not only relaxes existing constraints, but also unveils new, otherwise inaccessible regions of parameter space. Crucially, it does not merely serve as a loophole to evade limits, it provides a fundamentally different cosmological framework that enhances the discovery potential of DM models.

In this context, a DM signal consistent with the detection of a $Z'$ boson could offer indirect evidence for a non-standard cosmological era preceding BBN. For example, the benchmark point $m_{Z'} \sim 5$~GeV (or equivalently $m_\chi \sim 2.45$~GeV) with a gauge coupling $\gmt \sim 6 \times 10^{-7}$ cannot be explained within the standard cosmological picture, but emerges naturally in scenarios with an EMD phase. This suggests that future collider experiments could, in principle, provide indirect access to the epoch between inflationary reheating and BBN.

\section{Conclusions}\label{sec:conclusions}
We have performed a detailed and extended analysis of the phenomenology of a vector-like Dirac fermion dark matter (DM) candidate within the framework of the $L_\mu - L_\tau$ model, considering scenarios beyond the standard freeze-out mechanism.

The presence of an early matter-dominated (EMD) era introduces non-standard cosmological dynamics that significantly alter the Hubble expansion rate in the early universe. This, in turn, affects the thermal history and the DM freeze-out process. Our results demonstrate that in Phases I to III of the evolution of the EMD, it is possible to achieve the observed relic abundance even in regions of parameter space far from the resonance, such as $m_\chi / m_{Z'} = 0.3$. Notably, these regions can remain viable under current experimental constraints for a large range of DM masses. 

Moreover, when DM production occurs near the resonance ($m_{Z'} \sim 2\, m_\chi$), the EMD scenario opens up entirely new viable regions in the parameter space that would otherwise be excluded or untestable in the standard cosmology regime. This highlights a key advantage of the EMD framework: it enables thermal DM masses at different scales. Consequently, if a future detection reveals a DM signal consistent with a $Z'$ boson that lies within such regions, it may serve as an indirect probe of the pre-BBN universe.

In conclusion, non-standard cosmological scenarios, particularly EMD, play a dual role in the DM phenomenology. First, they help to evade current experimental bounds that exclude large portions of the standard freeze-out parameter space. Second, they unlock new viable regions, especially near resonance, where the standard mechanism tends to saturate. Thus, incorporating EMD cosmology provides not only an extended theoretical landscape but also a meaningful connection to the early universe's thermal history.

\acknowledgments
The authors express special thanks to the Mainz Institute for Theoretical Physics (MITP) of the Cluster of Excellence PRISMA+ (Project ID 390831469) for its hospitality and support. The authors also thank Manfred Lindner and Juri Smirnov for discussions. NB received funding from the grants PID2023-151418NB-I00 funded by MCIU/AEI/10.13039/501100011033/ FEDER and PID2022-139841NB-I00 by MICIU/AEI/10.13039/501100011033 and FEDER, UE. JPN acknowledges support from the Programa Institucional de Internacionalização (PrInt) and the Coordenação de Aperfeiçoamento de Pessoal de Nível Superior (CAPES) under the CAPES-PrInt Grant No. 88887.912033/2023-00. JPN also thanks the University of Liverpool for the hospitality during the final stages of this project. JS acknowledges funding by the {\it Dirección de Gestión de la Investigación} at PUCP, through grant DFI-PUCP-PI1144. FSQ is a Simons Foundation grantee (Award Number:1023171-RC), acknowledges support from CNPQ Grants 403521/2024-6, 408295/2021-0, 403521/2024-6, 406919/2025-9, 351851/2025-9, the FAPESP Grants 2021/01089-1, 2023/01197-4, ICTP-SAIFR 2021/14335-0, and the ANID-Millennium Science Initiative Program ICN2019\_044. This work is partially funded by FINEP under project 213/2024 and was carried out in part through the IIP cluster {\it bulletcluster}. The authors thank Hans Thomas Janka and Daniel Kresse for sharing the radial profiles simulation data, and the Garching Core-Collapse Supernova Archive (\href{https://wwwmpa.mpa-garching.mpg.de/ccsnarchive/}{https://wwwmpa.mpa-garching.mpg.de/ccsnarchive/}).

\appendix
\section{Experimental constraints} \label{app:current}
In this appendix, we discuss several current bounds and future probes of the model. They are summarized in Fig.~\ref{fig:ZpBounds}.

\subsection{Current constraints}
\subsubsection{Effective number of neutrino species}
\begin{figure*}[!ht]
    \centering
    \includegraphics[width=.49\linewidth]{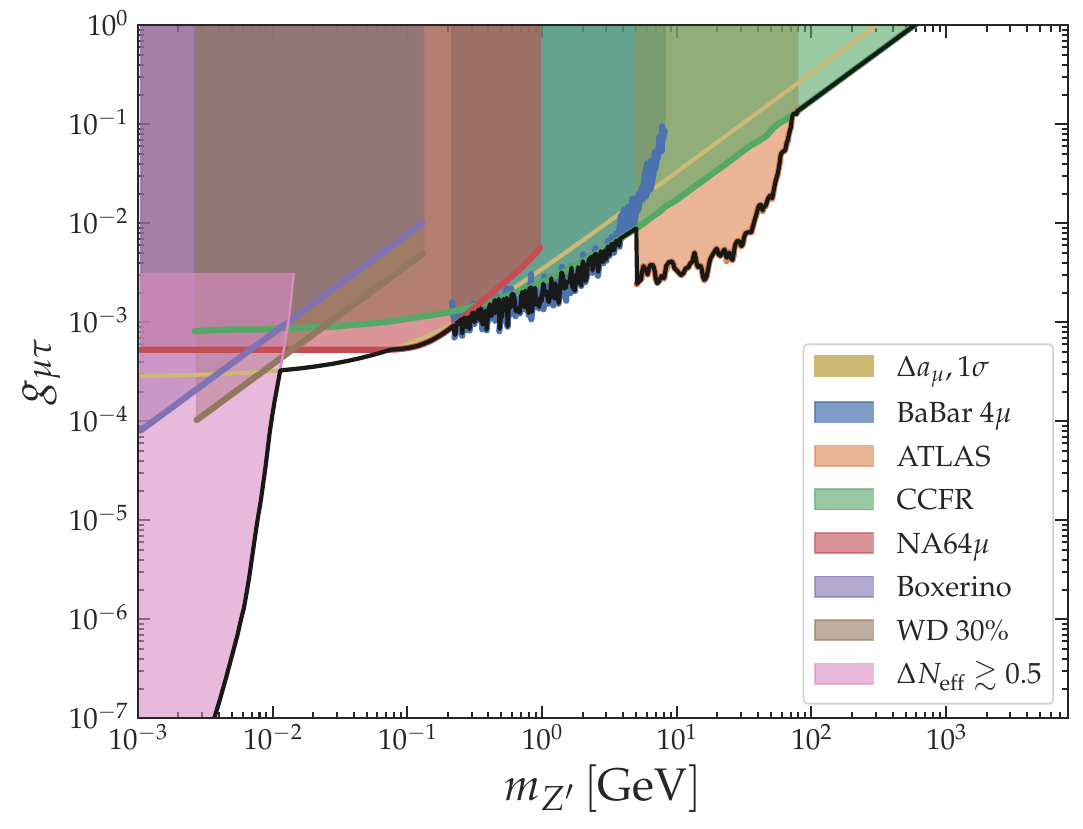}
    \includegraphics[width=.49\linewidth]{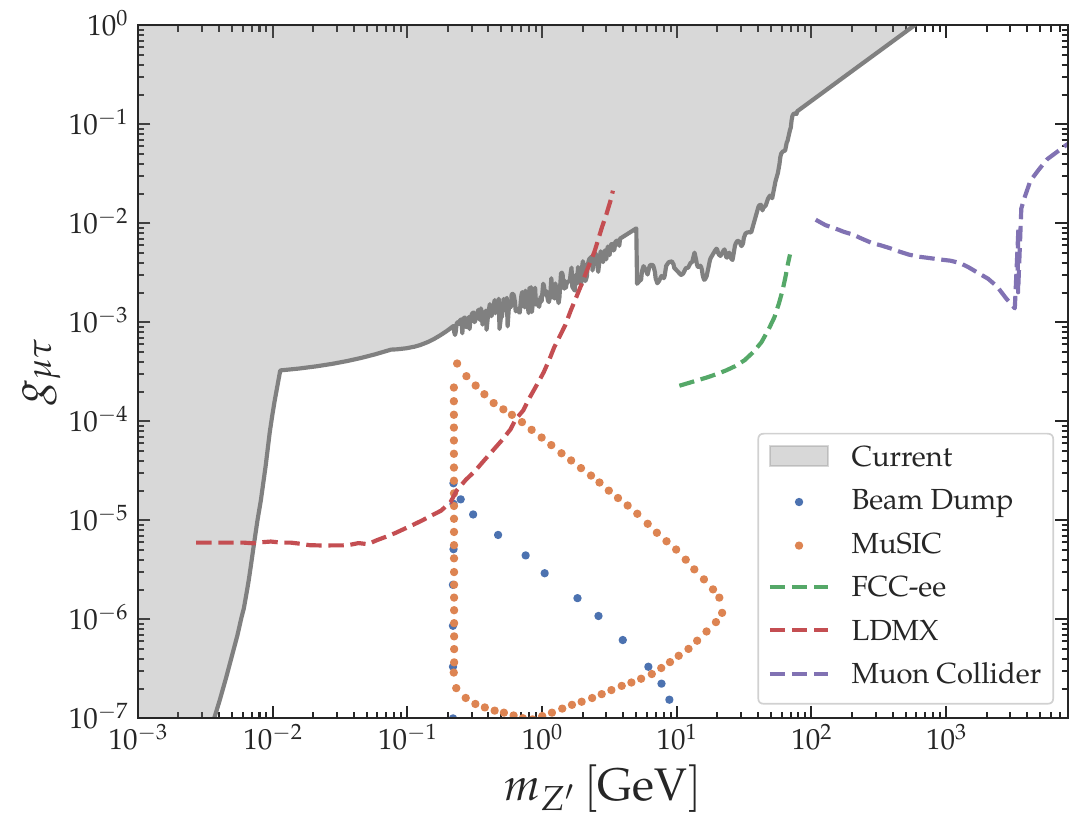}
    \caption{Current constraints (left panel) and future sensitivities (right panel) over the $Z'$ boson parameter space. These bounds represent experimental and observational efforts to probe this leptophilic fundamental interactions.}
    \label{fig:ZpBounds}
\end{figure*}
Given the mass and gauge coupling ranges we are considering, the decay of $Z'$ and the DM annihilation contribute to the total radiation density before matter-radiation equality (recombination), consequently increasing the effective number of neutrino species after neutrino decoupling, around $T_{\rm dec} \simeq 2$~MeV. Assuming only neutrinos and photons are present, we can obtain such a contribution from~\cite{Escudero:2019gzq, Holst:2021lzm}
\begin{equation}
    \Delta N_{\rm eff}  \equiv \frac{8}{7}\left(\frac{11}{4} \right)^{3/4} \frac{\delta \rho_\nu}{\rho_\gamma}\Bigg|_{T = T_{\rm dec}}\,,
\end{equation}
where $\delta \rho_\nu$ accounts for the extra amount of neutrino energy density coming from processes such as $\bar\chi \chi \to \bar\nu \nu \, /\, e^- e^+$ and $Z' \to \bar\nu\nu\,/\,e^- e^+$ after the inverse processes are allowed kinematically. The extra energy density can be obtained assuming that $\chi$ and $Z'$ were in thermal equilibrium with the SM bath at early times. The SM predicts $\Delta N_{\rm eff} =  3.045$~\cite{Mangano:2005cc, deSalas:2016ztq}, which is consistent with Planck measurements, $\Delta N_{\rm eff} = 2.99 \pm 0.17$~\cite{Planck:2018vyg}. Imposing $\Delta N_{\rm eff} \lesssim 0.5$, for $\gmt \gtrsim 4 \times 10^{-9}$, the $Z'$ mass obeys $m_{Z'} \lesssim 0.01$~GeV~\cite{Escudero:2019gzq}, as shown in the left panel of Fig.~\ref{fig:ZpBounds}.

\subsubsection{Hot white dwarf cooling}
A hot white dwarf (WD) star primarily loses energy through plasmon decay into neutrinos in its interior, resulting in an overall cooling effect. Ref.~\cite{Foldenauer:2024cdp} calculated the neutrino emissivity associated with this process, incorporating contributions from the $Z'$ boson in the $\lmlt$ model, and derived constraints on the sub-GeV parameter space. Their analysis considered a WD with mass $M_{\rm WD} = 1\, M_{\odot}$ and temperature $T_{\rm WD} = 10^8$~K. The constraints were obtained by estimating the relative excess luminosity induced by the $Z'$ contribution with respect to the SM prediction
\begin{equation}
    \varepsilon^{\rm BSM} \equiv \frac{L_{\rm WD}^{\rm SM+BSM} - L_{\rm WD}^{\rm SM}}{L_{\rm WD}^{\rm SM}}\,,
\end{equation}
where the luminosities $L_{\rm WD}^{\rm SM}$ and $L_{\rm WD}^{\rm SM+BSM}$ are computed by integrating the total neutrino emissivity $\mathcal{W}_{\rm tot}$ over the radius of the WD, $R_{\rm WD}$, assuming spherical symmetry 
\begin{equation}
    L_{\rm WD} = 4\pi \int_{0}^{R_{\rm WD}} dr\, r^2\, \mathcal{W}_{\rm tot}\,.
\end{equation}
Based on the current best-fit values of the hot WD neutrino luminosity function, which allows for a variation in cooling relative to the SM within the range $0.66 \lesssim f_s \lesssim 1.31$, they excluded the brown region shown in Fig.~\ref{fig:ZpBounds} for an additional cooling contribution of $\varepsilon^{\rm BSM} = 0.3$ (that is, 30\%) at 90\% CL.

Another interesting astrophysical limit might arise from old neutron stars. The authors in Ref.~\cite{Bell:2025acg} derived projected limits from these neutron stars, and predict that a substantially large yet unexplored region of the DM mass, $10^{-1}\,\,{\rm GeV} \leq m_\chi \leq 10^2$~GeV, could be probed for couplings $g_{\mu\tau} \sim 10^{-3}$ and $Q_{\mu\tau}^\chi \gtrsim 10^{-3}$.

\subsubsection{The muon anomalous magnetic moment}
\begin{figure}
    \centering
    \includegraphics[width=\linewidth]{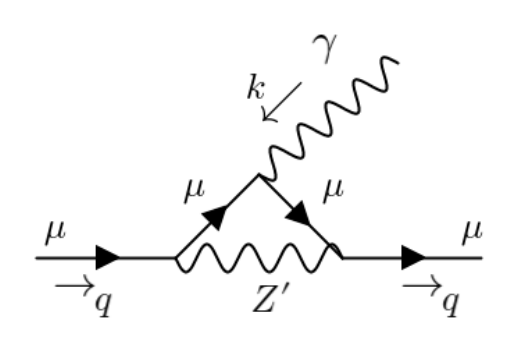}
    \caption{The Feynman diagram of the $Z'$ boson contribution to the muon anomalous magnetic moment.}
    \label{fig:g2muon}
\end{figure}

In the vanilla $\lmlt$ model, the contribution of $Z'$ to the muon magnetic moment at the one-loop level comes from the Feynman diagram in Fig.~\ref{fig:g2muon}, and the loop integral is written as~\cite{Baek:2001kca, Queiroz:2014zfa, Lindner:2016bgg}
\begin{equation}
    \Delta a_\mu = \frac{\gmt^2}{8\pi^2} \lambda^2 \int_0^1 dx \frac{2x^2(1-x)}{(1-x)(1-\lambda^2 x)+ \lambda^2 x}\,,
\end{equation}
where $\lambda \equiv m_\mu / m_{Z'}$. The lattice QCD community has significantly reduced the uncertainty in the contribution of the leading-order hadronic vacuum polarization to $\Delta a_\mu$ by employing a hybrid strategy that combines lattice QCD computations with data-driven inputs~\cite{Boccaletti:2024guq, Davies:2025pmx}. Additionally, the Muon $g-2$ Collaboration at Fermilab has recently reported its final result for the muon anomalous magnetic moment, achieving improved precision. The updated combined experimental average now reads $a_\mu ({\rm exp}) = 116 \, 592 \, 0715(145) \times 10^{-12}$~\cite{Muong-2:2025xyk}. Recent calculations indicate that the SM prediction for $a_\mu$ is now consistent with the updated experimental value within 1$\sigma$ uncertainty, $\Delta a_\mu = a^{\rm exp}_\mu (\rm new) - a^{\rm SM}_\mu (\rm new) = (39 \pm 64) \times 10^{-11}$~\cite{Aliberti:2025beg}. Although this may reduce the parameter space for new physics explanations of the muon $g-2$ anomaly, the $\lmlt$ model remains compelling. 

\subsubsection{Borexino}
$^7$Be neutrinos produced in the Sun scatter off electrons in the Borexino liquid scintillator experiment, providing a sensitive probe of BSM neutrino interactions with matter. In this context, the collaboration placed constraints on the $U(1)_{B-L}$ gauge interaction in Ref.~\cite{Bellini:2011rx}. These bounds were later rescaled to derive limits on the $\gmt$ coupling~\cite{Kaneta:2016uyt},\footnote{A similar procedure can be adopted to constrain other lepton-flavor gauge models~\cite{Bauer:2018onh}.} while in Ref.~\cite{Kamada:2018zxi}, the authors imposed an 8\% deviation from the SM neutrino-electron scattering rate to set bounds on the $Z'$ current.

More recently, the Borexino constraints were refined, requiring that the contribution of $Z'$ not exceed 23\% of the total rate at 90\% CL, and relaxed the previous limits~\cite{Gninenko:2020xys}. In Fig.~\ref{fig:ZpBounds}, we display the updated Borexino bounds in magenta, which are largely superseded by the WD cooling limits~\cite{Foldenauer:2024cdp}. Furthermore, the constraints from $\Delta N_{\rm eff}$ surpass the Borexino limits at lower $Z'$ masses, where the WD cooling limits are no longer effective.

\subsubsection{Neutrino trident production} 
When a muon neutrino scatters in the Coulomb field of a heavy nucleus, it can produce a pair $\mu^+\, \mu^-$. This process is known as neutrino trident production, $\nu_\mu\, N \to \nu_\mu\, N\, \mu^+\, \mu^-$~\cite{Altmannshofer:2016jzy}. This rare process has been measured by several experiments in good agreement with the SM predictions. In particular, the CHARM-II collaboration~\cite{CHARM-II:1990dvf} and the CCFR collaboration~\cite{CCFR:1991lpl} reported
\begin{align}
    \sigma_{\rm CHARM-II} /\sigma_{\rm SM} &= 1.58 \pm 0.57 \,,\\
    \sigma_{\rm CCFR} /\sigma_{\rm SM} &= 0.82 \pm 0.28\,.
\end{align}
Although Ref.~\cite{Altmannshofer:2016jzy} provides a simplified analytical bound on the $Z'$ mass as a function of the $\gmt$ gauge coupling, this result is valid only within a specific regime. In our work, we instead adopt the CCFR limits from Ref.~\cite{Altmannshofer:2014pba}, which performed a full numerical evaluation over the relevant parameter space and excluded the green region in Fig.~\ref{fig:ZpBounds} at 95\% CL.

\subsubsection{ATLAS}
The ATLAS collaboration has reported constraints on the $Z'$ gauge boson via searches in both neutral-current and charged-current Drell–Yan (DY) processes, where a $Z'$ can be emitted through final-state radiation of a $\mu^+\,\mu^-$ pair. These limits apply within the mass range 5~GeV $\lesssim m_{Z'} \lesssim$ 81~GeV. We discuss both cases separately below.

The neutral-current DY search was presented in Ref.~\cite{ATLAS:2023vxg}, targeting $4\mu$ final states with a center-of-mass energy of $\sqrt{s} = 13$~TeV and an integrated luminosity of 139~fb$^{-1}$. In this process, a virtual $Z$ boson produced in proton-proton collisions decays into a $\mu^+\,\mu^-$ pair, one of which radiates a $Z'$ boson. Subsequently, the $Z'$ decays into an additional $\mu^+\,\mu^-$ pair. The collaboration established upper bounds at 95\%~CL on the production cross-section times branching ratio for the process $p\,p \to Z^{(\star)} \to Z'\,\mu^+\,\mu^- \to \mu^+\,\mu^-\,\mu^+\,\mu^-$. 

A complementary charged-current DY search was reported in Ref.~\cite{ATLAS:2024uvu}, targeting $3\mu$ final states with a large missing transverse energy at $\sqrt{s} = 13$~TeV and an integrated luminosity of 140~fb$^{-1}$. In this scenario, a virtual $W^\pm$ boson decays into a $\mu^\pm$ and a $\nu_\mu$, either of which can emit an on-shell $Z'$ boson decaying into a pair $\mu^+\,\mu^-$. Using a similar analysis strategy, the collaboration placed upper bounds at 95\%~CL on the cross-section times branching ratio for the process $p\,p \to W^{\pm\,(\star)} \to Z'\,\mu^{\pm}\,\nu_\mu \to \mu^+\,\mu^-\,\mu^{\pm}\,\nu_\mu$.  

Furthermore, Ref.~\cite{ATLAS:2024uvu} presents a combined analysis of both searches, resulting in stronger exclusion limits. These are displayed in orange in Fig.\ref{fig:ZpBounds}.

\subsubsection{NA64\texorpdfstring{$\mu$}{mu}}
NA64$\mu$ is a fixed-target experiment that employs the missing energy-momentum technique to search for dark-sector particles. In this setup, a muon beam from the M2 beamline at the CERN Super Proton Synchrotron (SPS) scatters off atomic nuclei in a target via a bremsstrahlung-like process, producing a $Z'$ boson that promptly decays into invisible particles: $\mu\,N \to \mu\,N\,Z'$, followed by $Z' \to {\rm invisible}$~\cite{NA64:2024klw}. In the context of the $\lmlt$ model, the invisible final states consist only of neutrinos and DM. Focusing on the DM channel, the collaboration has placed stringent limits for $m_\chi < 1$~GeV, corresponding to the red region shown in Fig.~\ref{fig:ZpBounds}. Although not explicitly shown here, the NA64$e$ experiment, based on an electron beam, sets even stronger exclusion limits in the low-mass region, particularly for DM masses below 1~MeV\cite{Andreev:2024lps}.
\subsubsection{BaBar \texorpdfstring{4$\mu$}{4-muon} \& Belle-II}
BaBar and Belle-II search for the $\lmlt$ gauge boson through the process $e^+\,e^- \to \mu^+\,\mu^-\,Z'$, with subsequent decay $Z' \to \mu^+\,\mu^-$~\cite{BaBar:2016sci, Belle-II:2024wtd}. In this scenario, the $Z'$ boson is produced as final-state radiation from one of the muons. The analyses focus on events containing two pairs of oppositely charged tracks, corresponding to $4\mu$ final states. The signal manifests itself as a narrow resonance in the invariant mass distribution of muon pairs. Although both experiments yield similar constraints, BaBar remains slightly more sensitive, which is expressed in the blue region in Fig.~\ref{fig:ZpBounds}.

\subsection{Future probes}
Now, we present proposed experimental setups that could probe the present scenario.

\subsubsection{Light Dark Matter eXperiment (LDMX) \texorpdfstring{$M^3$}{M3}}
The LDMX experiment searches for missing momentum in high-luminosity electron–nucleus fixed-target collisions~\cite{Berlin:2018bsc}. The energy of the electron beam lies in the range of 4–16~GeV, and the experimental setup is conceptually similar to that of NA64. However, LDMX takes advantage of the higher intensities available from a primary electron beam, enabling it to probe smaller couplings. Additionally, the precise measurement of the recoil electron momentum helps to discriminate between signal and background events.

A complementary proposal involves replacing the electron beam with a muon beam, resulting in the so-called LDMX M$^3$ configuration~\cite{Kahn:2018cqs}. Although the muon-beam setup is constrained by lower available beam intensities, this limitation can be partially offset by employing a thicker target. Under this configuration, the experiment can probe the parameter space of muon-philic $Z'$ gauge bosons via their invisible decays, yielding competitive sensitivity curves. In Fig.~\ref{fig:ZpBounds}, right panel, the red dashed lines show the LDMX projected sensitivity.

\subsubsection{MuSIC and muon beam dump}
The Muon (Synchrotron) Ion Collider (MuSIC) is envisioned to be the successor to the Electron-Ion Collider (EIC)\cite{Accardi:2012qut, Aschenauer:2014cki}, which is currently under construction at Brookhaven National Laboratory. In Ref.~\cite{Davoudiasl:2024fiz}, the authors presented the projected exclusion limits for MuSIC and compared them with those of the proposed Muon Beam Dump experiment~\cite{Cesarotti:2023sje}.

MuSIC aims to investigate the muon bremsstrahlung process $\mu\,{\rm Au} \to \mu\,{\rm Au}\,Z'$, where the $Z'$ boson is produced by coherent scattering off gold ions. The experimental setup involves colliding a high-energy muon beam with a beam of gold nuclei. The muon beam operates at a center-of-mass energy of $\sqrt{s} = 1$~TeV, while the gold ions are accelerated to $100$~GeV per nucleon, producing a total center-of-mass energy of $\sqrt{s} = 8.9$~TeV. The integrated luminosity over five years of operation is estimated as $L_I = 400\,{\rm fb}^{-1}/A$, where $A$ is the atomic mass number. For gold, with $A = 196.97$, they assumed it to be $L_I \simeq 2$~fb$^{-1}$.

In contrast, the Muon Beam Dump experiment employs a fixed lead target and a $1.5$~TeV muon beam, delivering $10^{20}$ muons on target and thus achieving significantly higher effective luminosity. Despite the different experimental configurations, both setups are sensitive to muon-philic $Z'$ bosons, assuming background-free conditions and focusing on the region above the dimuon threshold. Although the Muon Beam Dump experiment offers greater sensitivity to small couplings, MuSIC has the advantage of accessing larger $Z'$ masses due to its broader kinematic reach.

The projected sensitivities, at 95\% CL, of MuSIC and Muon Beam Dumb experiments are shown as orange and blue dots in Fig.~\ref{fig:ZpBounds}, respectively.

\subsubsection{Muon collider}
A future TeV-scale muon collider would offer a powerful probe of muon-philic $Z'$ bosons with masses above $100$~GeV. In Ref.~\cite{Huang:2021nkl}, the authors assumed a center-of-mass energy of $\sqrt{s} = 3$~TeV and an integrated luminosity of $L = 1$~ab$^{-1}$; the analysis considers SM backgrounds from the processes $\mu^+ \,\mu^- \to \bar{f} \,f \,\gamma$ and $\mu^+ \, \mu^- \to \bar{f} \, f$, where $f = \mu, \tau, \nu_\mu$, or $\nu_\tau$. The projected sensitivity curve, shown as a dashed magenta line in Fig.~\ref{fig:ZpBounds}, is comparable to that of ATLAS but extends to significantly higher $Z'$ masses. Therefore, a high-energy muon collider would serve as a complementary probe of new muon-philic interactions at the energy frontier. Note that there is an ongoing effort to build a high-energy muon collider up to $\sqrt{s} = 10$~TeV~\cite{InternationalMuonCollider:2025sys}.

\subsubsection{FCC-ee}
This will be an $e^+\,e^-$ experiment in the Future Circular Collider (FCC).  In Ref.~\cite{Airen:2024iiy}, the authors investigated the $Z$-pole operation mode at the Future Circular Collider (FCC) as a benchmark to probe the $Z'$ boson associated with the $\lmlt$ gauge symmetry.\footnote{The study also considered the dark photon scenario, including the $ZH$ operation mode.} In this setup, $Z$ bosons are produced via the process $e^+ \, e^- \to Z$, and the $Z'$ arises from exotic $Z$ decays, specifically through $Z \to \mu^+ \, \mu^- \, Z'$. The visible and invisible decay channels of $Z'$ were analyzed, and 95\% CL projected sensitivities were derived for $Z'$ masses ranging from $10$~GeV to values close to the mass of the $Z$ boson. In Fig.~\ref{fig:ZpBounds}, the dashed green line represents the projected sensitivity for the visible decay channel, which yields stronger limits.\\

\bibliographystyle{JHEPfixed}
\bibliography{references}

\end{document}